\documentclass[]{elsart}
\usepackage{amsmath}
\usepackage{amssymb}
\usepackage{graphicx}

\def\issue(#1,#2,#3){#1 (#3) #2} 

\def\opcit(#1){ {\em op. cit.}, #1}

\def\ARNPS(#1,#2,#3){Ann.\ Rev.\ Nucl.\ Part.\ Sci.\ \issue(#1,#2,#3)}
\def\CPC(#1,#2,#3){Comp.\ Phys.\ Comm.\ \issue(#1,#2,#3)}
\def\CIP(#1,#2,#3){Comput.\ Phys.\ \issue(#1,#2,#3)}
\def\EPJC(#1,#2,#3){Eur.\ Phys.\ J.\ C\ \issue(#1,#2,#3)}
\def\EPJA(#1,#2,#3){Eur.\ Phys.\ J.\ A\ \issue(#1,#2,#3)}
\def\IEEETNS(#1,#2,#3){IEEE Trans.\ Nucl.\ Sci.\ \issue(#1,#2,#3)}
\def\NP(#1,#2,#3){Nucl.\ Phys.\ \issue(#1,#2,#3)}
\def\NIM(#1,#2,#3){ Nucl.\ Instrum.\ and Meth.\ \issue(#1,#2,#3)}
\def\PL(#1,#2,#3){Phys.\ Lett.\ \issue(#1,#2,#3)}
\def\PRD(#1,#2,#3){Phys.\ Rev.\ D \issue(#1,#2,#3)}
\def\PRL(#1,#2,#3){Phys.\ Rev.\ Lett.\ \issue(#1,#2,#3)}
\def\SJNP(#1,#2,#3){Sov.\ J. Nucl.\ Phys.\ \issue(#1,#2,#3)}
\def\ZPA(#1,#2,#3){Z.\ Phys.\ A \issue(#1,#2,#3)}
\def\PAN(#1,#2,#3){Phys.\ Atom.\ Nucl.\ \issue(#1,#2,#3)}
\def\FBS(#1,#2,#3){Few\ Body\ Syst.\ \issue(#1,#2,#3)}
\def\JPCS(#1,#2,#3){J.\ Phys.\ Conf.\ Ser.\ \issue(#1,#2,#3)}
\def\JPG(#1,#2,#3){J.\ Phys.\ G\ \issue(#1,#2,#3)}
\def\NPPS(#1,#2,#3){Nucl.\ Phys.\ Proc.\ Suppl.\ \issue(#1,#2,#3)}
\def\IJMPA(#1,#2,#3){Int.\ J.\ Mod.\ Phys.\ A\ \issue(#1,#2,#3)}
\begin{document}

\begin{frontmatter}

\title{Search for a pentaquark decaying to $p K_S^0$}

\collaboration{The~FOCUS~Collaboration}\footnotemark
\author[ucd]{J.~M.~Link}
\author[ucd]{P.~M.~Yager}
\author[cbpf]{J.~C.~Anjos}
\author[cbpf]{I.~Bediaga}
\author[cbpf]{C.~Castromonte}
\author[cbpf]{A.~A.~Machado}
\author[cbpf]{J.~Magnin}
\author[cbpf]{A.~Massafferri}
\author[cbpf]{J.~M.~de~Miranda}
\author[cbpf]{I.~M.~Pepe}
% ...

\address[ucd]{University of California, Davis, CA 95616}
\address[cbpf]{Centro Brasileiro de Pesquisas F\'\i sicas, Rio de Janeiro, RJ, Brazil}
% ...

\footnotetext{See \textrm{http://www-focus.fnal.gov/authors.html} for additional author information.}

\begin{abstract}
We present a search for a pentaquark decaying strongly to $pK_S^0$ in $\gamma N$ 
collisions at a center-of-mass energy up to 25~GeV/$c^2$.  Finding no
evidence for such a state in the mass range of 1470~MeV/$c^2$ to 2200~MeV/$c^2$, 
we set limits on the yield and on the cross section 
times branching ratio
relative to $\Sigma^*(1385)^\pm$ and $K^*(892)^+$.
\end{abstract}


\end{frontmatter}

\section{Introduction}

Nearly 30 years ago Jaffe proposed the existence of bound 
(mass below threshold for strong decay) multiquark states including
$Q\overline{Q}q\overline{q}$ states and the $H$ dihyperon~\cite{jaffe} based on calculations
using the bag model~\cite{bag}.  As the years passed and no convincing evidence for non mesonic
and non baryonic states was found the field languished.  

Between January 2003 and March 2004, however, the pentaquark field was reenergized when no less than 
ten independent pentaquark observations at a mass around $1540\;\textrm{MeV}\!/c^2$ were 
reported~\cite{leps,diana,clas1,saphir,asratyan,clas2,hermes,svd,cosytof,zeus}.  The presumed quark content
of the reported states was $(\overline{s}uudd)$.  The reported widths were consistent with detector resolution
even though a strong decay is allowed.  Even more amazing was the prediction six years before by 
Diakonov \textit{et al.} of just such a state at a mass around $1530\;\textrm{MeV}\!/c^2$ and a width less
than $15\;\textrm{MeV}/c^2$~\cite{diakonov}.  Observations of a doubly-strange pentaquark~\cite{na49} and a 
charm pentaquark~\cite{h1} have also been reported by single experiments.

The original observations were made in the $nK^+$ mode.  For a state to decay 
strongly to $nK^+$, it must be composed of at least 5 quarks.  Other observations have been made in the 
$pK_S^0$ mode which is not manifestly exotic since the $K_S^0$ can originate from a $K^0$ or 
$\overline{K}{}^0$; a $p\overline{K}{}^0$ decay is not exotic while a $pK^0$ decay is exotic.
Since that time, many other experiments have failed to find evidence of 
pentaquarks~\cite{aleph,babar,belle,bes,cdf,herab,hypercp,napolitano,lep,phenix,sphinx}.
Most of these experiments are higher statistics and higher energy than the observing 
experiments and generally search the $pK_S^0$ decay mode.  Recently CLAS, which previously reported
two observations failed to find pentaquarks in a third attempt~\cite{clasno}.
This letter presents a search of the FOCUS data for the $\Theta^+(\overline{s}uudd)$ 
pentaquark candidate in the decay mode $\Theta^+\to pK_S^0$.  Cross sections will be measured 
relative to three well known states with a similar decay topology: $\Sigma^*(1385)^\pm \to \Lambda^0\pi^\pm$ 
(two states) and 
$K^*(892)^+\to K_S^0 \pi^+$~\footnote{Charged conjugate states are implied unless explicitly stated otherwise}.

\section{Event reconstruction and selection}

The FOCUS experiment recorded data during the 1996--7 fixed-target run at Fermilab.  
A photon beam obtained from bremsstrahlung of 300~GeV electrons and positrons impinged
on a set of BeO targets.  Four sets of silicon strip detectors, each with three views, 
were located downstream of the targets for vertexing and track finding.  For most of 
the run, two pairs of silicon strips were also interleaved with the target segments for more
precise vertexing~\cite{tsilicon}.  Charged particles were tracked and momentum analyzed as they passed
through one or two dipole magnets and three to five sets of multiwire proportional chambers 
with four views each.  Three multicell threshold \v{C}erenkov counters, two electromagnetic
calorimeters, and two muon detectors provided particle identification.  A trigger which 
required, among other things, $\gtrsim$25~GeV of hadronic energy passed 6 billion events for 
reconstruction.

The data used for this analysis come from a subset of FOCUS data which contain vee candidates 
($K_S^0\to \pi^+\pi^-$ and 
$\Lambda^0\to p\pi^-$).  
There are four vee candidate types which are
used in this analysis.  SSD vees decay in the vertex region and have decay tracks which are 
found in the silicon system.  Magnet vees decay further downstream and tracks are reconstructed 
in the wire chambers.  The magnet vees are divided into three types SS, TS, TT for stub-stub, 
track-stub, and track-track depending on whether the decay particles are tracked in only the
upstream 3 wire chambers (stubs) or in all 5 wire chambers (tracks).  A full description of vee
reconstruction in FOCUS can be found in Ref.~\cite{vees}.

\begin{figure}
\centerline{\includegraphics[width=2.7in,height=2.3in]{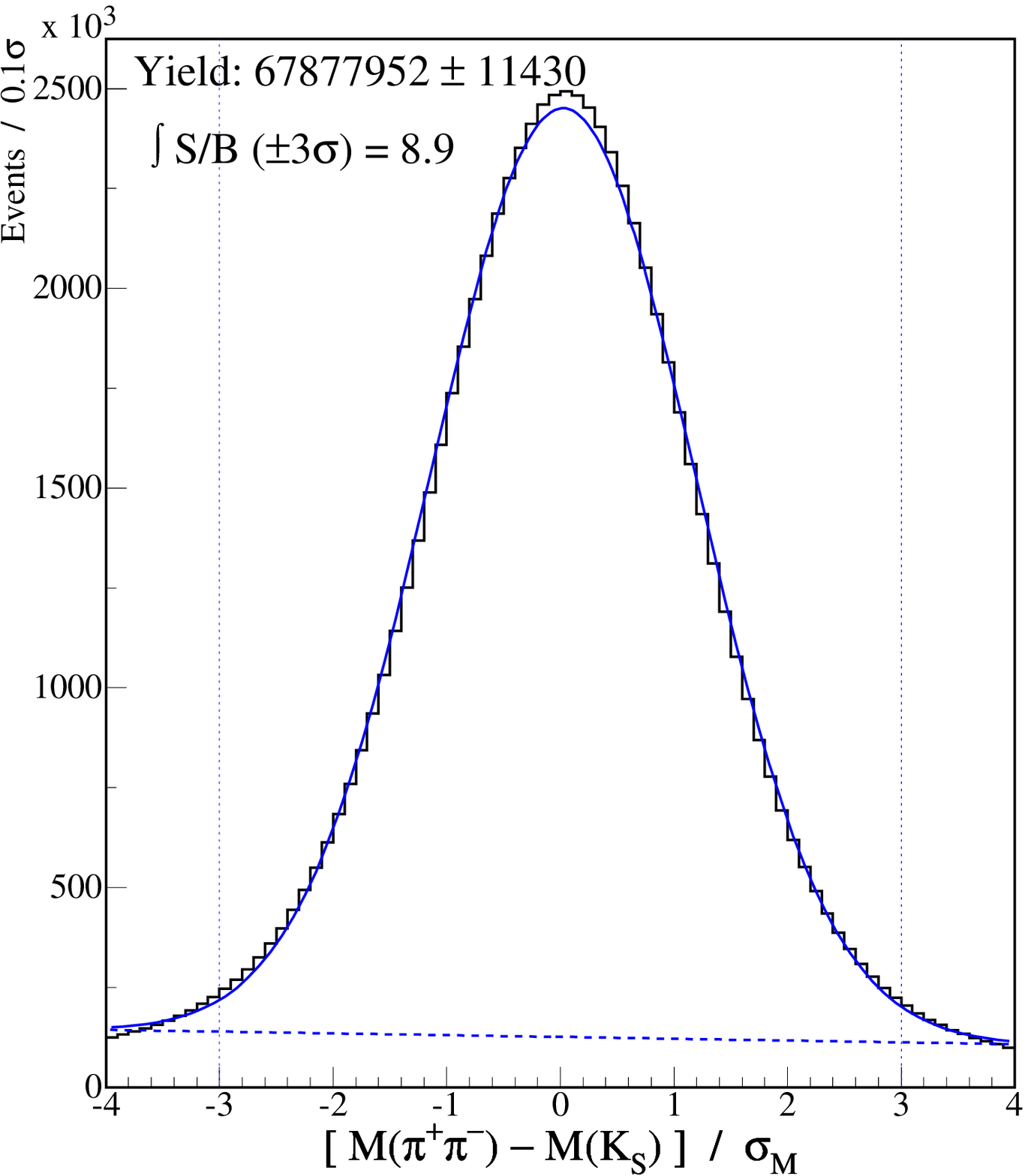}\hspace{-3pt}
\includegraphics[width=2.7in,height=2.3in]{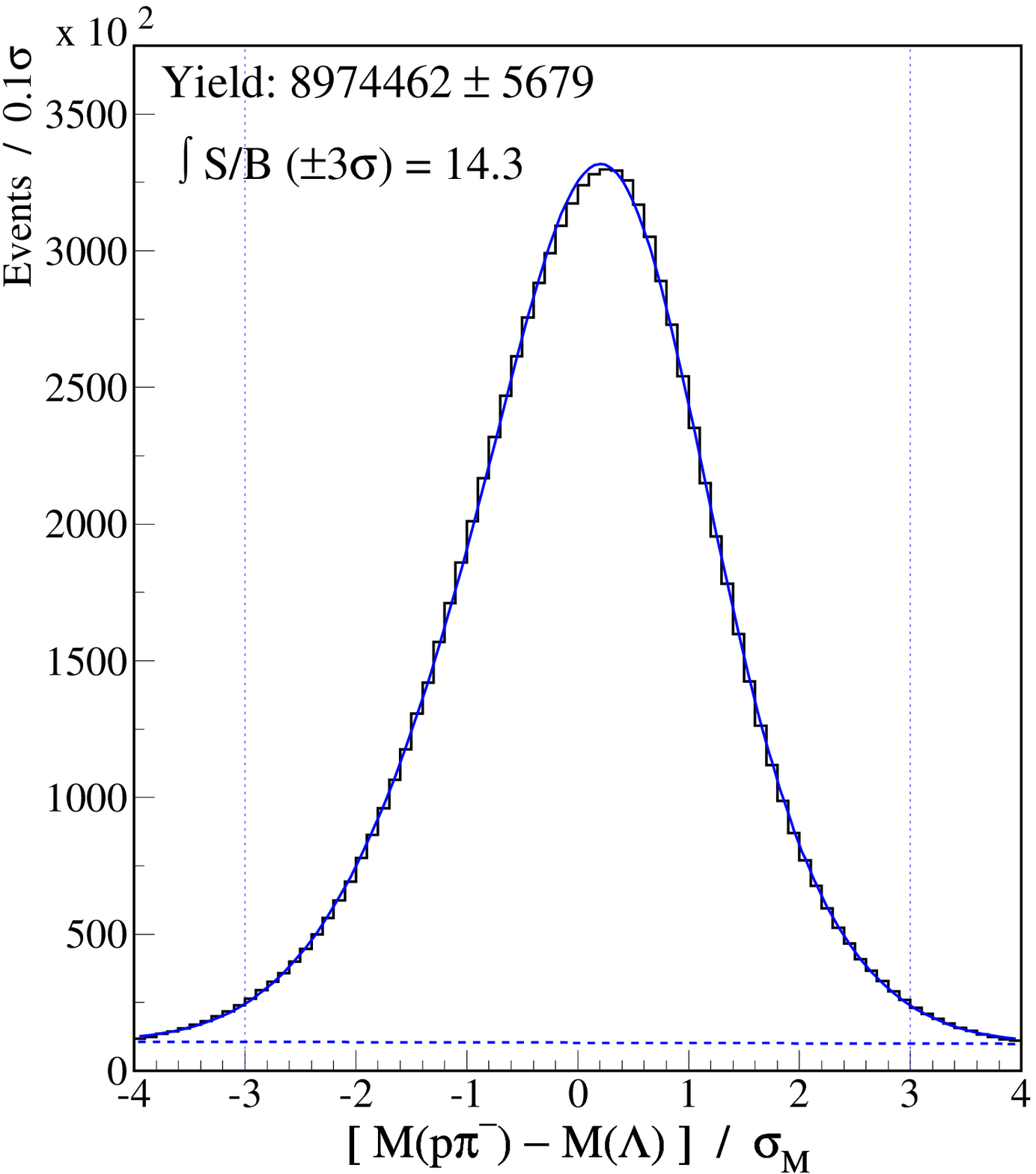}}
\caption{The normalized mass plots of $K_S^0\to\pi^+\pi^-$ and $\Lambda^0\to p^-\pi^-$ candidates.
Events inside the vertical lines are selected for analysis.}
\label{fig:vees}
\end{figure}

The vee selection requires a reconstructed vee vertex with some quality, particle
identification, and mass cuts.  The quality requirements include good track quality, 
a good vertex for the two tracks, and a minimum vee momentum of 5~GeV/$c$.
The mass requirement for a $K_S^0$ candidate is to have a normalized
mass within 4 (5) $\sigma$ of the nominal mass for SSD (magnet) vees.  The $\Lambda^0$ mass requirement
is for the invariant mass to be between 1.09~GeV/$c^2$ and 1.14~GeV/$c^2$.  
The particle identification cuts on the vee daughters use the FOCUS \v{C}erenkov identification 
algorithm~\cite{citadl}.   This algorithm returns negative log-likelihood 
(times two) values $\mathcal{W}_i(j)$ for track $j$ and hypothesis $i\in\left\{e,\pi,K,p\right\}$ 
based on the light yields in the phototubes covering the \v{C}erenkov cone of the track.  The information 
from all three \v{C}erenkov detectors is combined.
The vee daughter pion candidates must not be strongly inconsistent with the 
pion hypothesis: $\mathcal{W}_{\textrm{min}}(\pi) - \mathcal{W}_\pi(\pi) > -5$ where 
$\mathcal{W}_{\textrm{min}}$ is the minimum of the four hypotheses. 
The small phase space of the $\Lambda^0\to p\pi^-$ decay and the forward nature of the FOCUS spectrometer 
require the proton to have a higher lab momentum than the pion.  Thus, the higher momentum track is 
chosen to be the proton and it must have $\mathcal{W}_{\textrm{min}}(p) - \mathcal{W}_p(p) > -5$ 
and $\mathcal{W}_\pi(p)-\mathcal{W}_p(p) > 1$.  
To reduce combinatorics, the event is only kept if the number of vees passing the above cuts was no 
more than two.  At least one and no more than seven good quality charged tracks with momentum greater
than 5~GeV/$c$ must be found in addition to the vee(s).  These
tracks and any SSD vees must be consistent with originating from a single vertex with confidence level
greater than 1\%.  The total number of good quality vees plus good quality charged tracks must exceed two. 
After applying all of the above cuts, the vee sample within $3\sigma$ of the nominal mass is
selected which contains 
72 million $K_S^0\to \pi^+\pi^-$ (9 million $\Lambda^0\to p\pi^-$) candidates of which 90\% (95\%) are signal
as shown in Fig.~\ref{fig:vees}.  Each vee in an event is combined with the good quality charged tracks in 
the event to search for $K^*(892)^+ \to K_S^0\pi^+$,
$\Sigma^*(1385)^\pm\to \Lambda^0\pi^\pm$, and $\Theta^+\to pK_S^0$.  The \v{C}erenkov requirements on
the charged tracks are $\mathcal{W}_{\textrm{min}}(\pi)-\mathcal{W}_\pi(\pi) > -2$ for pions plus 
$\mathcal{W}_K(\pi) - \mathcal{W}_\pi(\pi) > 1$ for the pion from the $\Sigma^*(1385)^\pm$ decay.  The
proton from the $\Theta^+$ decay must have $\mathcal{W}_\pi(p) - \mathcal{W}_p(p) > 8$ and
$\mathcal{W}_K(p) - \mathcal{W}_p(p) > 3$.  The \v{C}erenkov cuts were optimized using signal Monte
Carlo with data background.  The proton requirements, in particular, are very stringent and reduce
the misidentification rate to nearly zero.

\begin{figure}
\centerline{\includegraphics[width=4.8in]{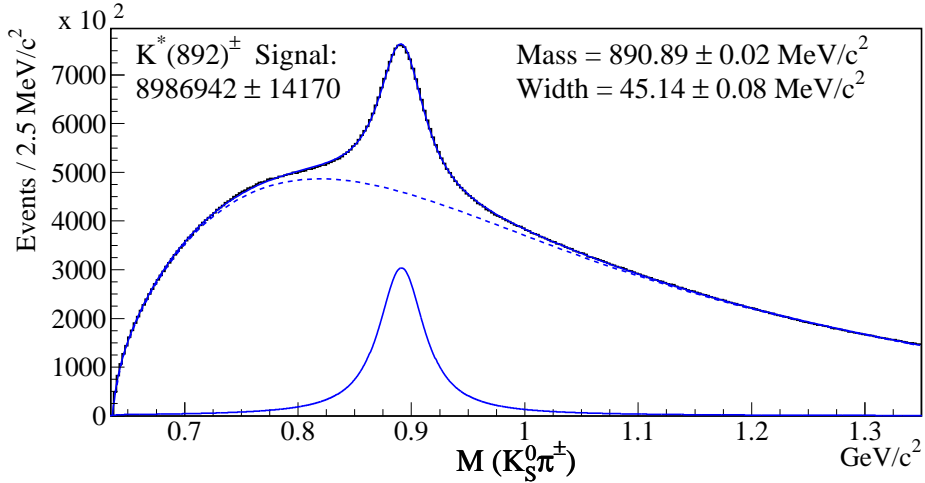}}
\caption[$K^*(892)^+$ fit with S-wave Breit-Wigner]{$K^*(892)^+$ fit with an S-wave Breit-Wigner and combinatorial background.}
\label{fig:kstar}
\end{figure}

\begin{figure}
\centerline{\includegraphics[width=5.5in]{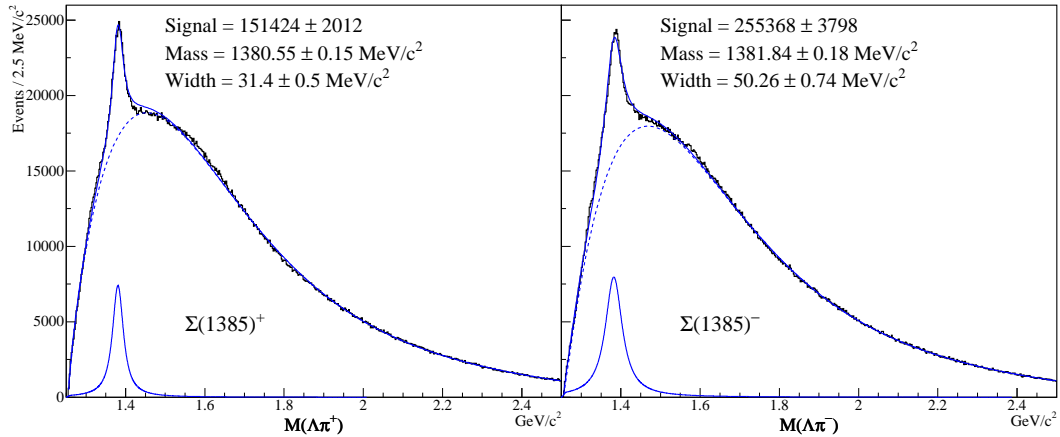}}
\caption[$\Sigma^*(1385)^\pm$ fits with S-wave Breit-Wigner]{$\Sigma^*(1385)^\pm$ fits with an S-wave 
Breit-Wigner and combinatorial background.}
\label{fig:sigma}
\end{figure}

The $K_S^0\pi^+$ and $\Lambda^0\pi^\pm$ are shown in Figs.~\ref{fig:kstar} and \ref{fig:sigma}, respectively.
In both cases, the signal was best fit with an S-wave Breit-Wigner with an energy independent width even 
though a P-wave energy dependent width would be more appropriate.  The Breit-Wigner 
was convoluted with a Gaussian for the detector resolution obtained from a Monte Carlo simulation.  The
$K^*(892)^+$, $\Sigma^*(1385)^+$, and $\Sigma^*(1385)^-$ resolutions are 5.1~MeV/$c^2$, 3.2~MeV/$c^2$, and
3.2~MeV/$c^2$.  The background was fit to the form $a q^b \exp{(cq + dq^2 + eq^3 + fq^4)}$ where 
$a$---$f$ are free parameters and $q$ is the $Q$-value (invariant mass minus component masses).

\section{Pentaquark search results}

\begin{figure}
\centerline{\includegraphics[width=5.0in]{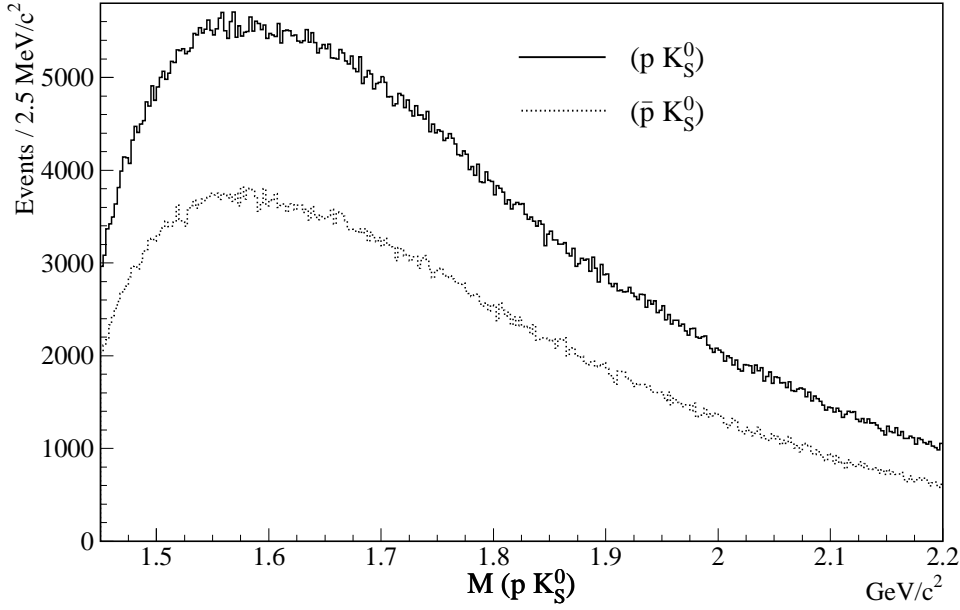}}
\caption{Invariant mass distribution of $p K_S^0$ separated by charge.
Standard cuts are applied.}
\label{fig:pks_plot_charge}
\end{figure}

\begin{figure}
\centerline{\includegraphics[width=5.0in]{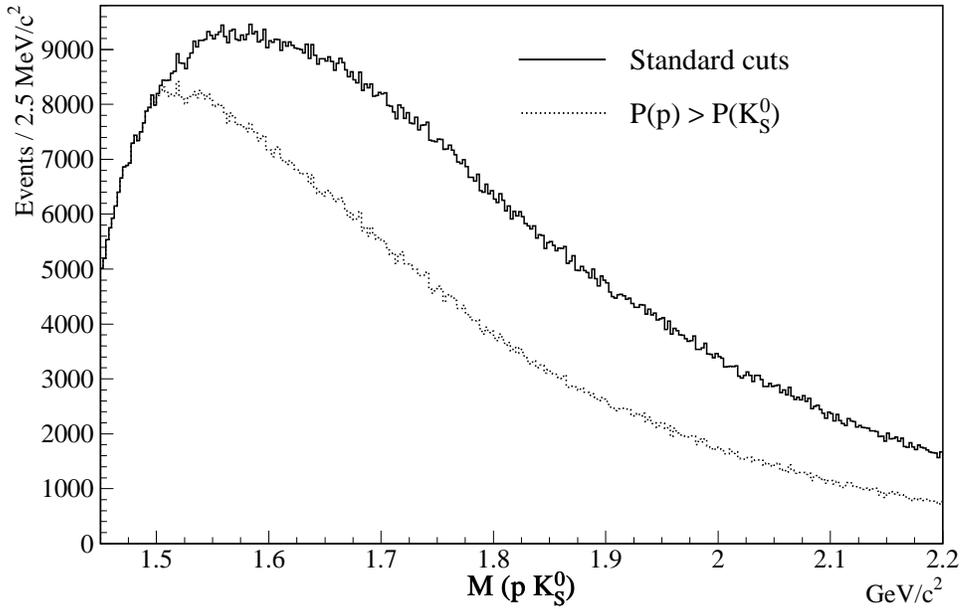}}
\caption{Invariant mass distribution of $p K_S^0$ for both charge states.
Solid histogram shows the result for standard cuts and the dashed histogram is with an additional cut requiring
the proton momentum be greater than the $K_S^0$ momentum.}
\label{fig:pks_plot_all}
\end{figure}

\begin{figure}
\centerline{\includegraphics[width=5.0in]{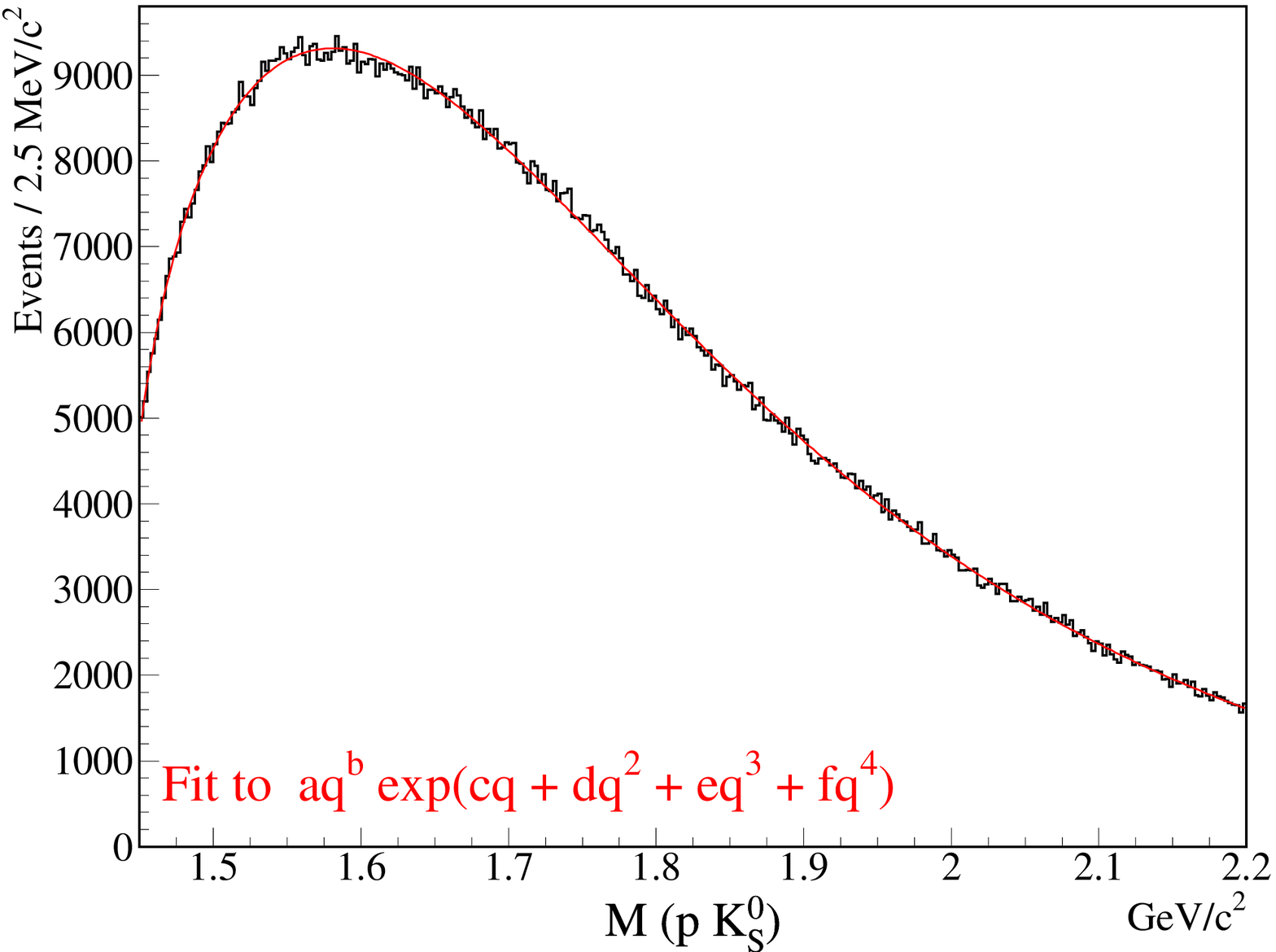}}
\caption{Invariant mass distribution of $p K_S^0$ for both charge states with standard cuts.}
\label{fig:pks_plot_fit}
\end{figure}

The $pK_S^0$ and $\overline{p}K_S^0$ invariant masses are plotted using the standard selection criteria in 
Fig.~\ref{fig:pks_plot_charge}.  There are no significant differences between the two charge states so for 
the remainder of the analysis we combine the charge conjugate states.
The combined sample with standard cuts and with an additional momentum
asymmetry cut is plotted in Fig.~\ref{fig:pks_plot_all}.  The momentum asymmetry cut, requiring the proton
to have a higher momentum than the $K_S^0$ in the pentaquark decay, has been suggested as a method of
reducing background.  The true effect of the cut is to sculpt the mass distribution into a more peaked structure
near the location of the previously observed pentaquarks and is not used in the analysis.
In Fig.~\ref{fig:pks_plot_fit} the total sample is fit to a background curve of the form 
$a q^b \exp{(cq + dq^2 + eq^3 + fq^4)}$ where $a$---$f$ are free parameters and $q$ is the $Q$-value: 
$q \equiv M(p K_S^0)-m_p-m_{K^0}$.
No evidence for a pentaquark near $1540$~MeV/$c^2$ or at any mass less than $2400$~MeV/$c^2$ is observed.
To set a limit on the yield we need to make some assumptions about the width of the state.  We consider two cases:
one with a natural width of 0 and one with a natural width of 15~MeV/$c^2$.  In the first case, the signal is
fit with a Gaussian with a width set by the experimental resolution.  In the second case, 
the signal is fit with an S-wave Breit-Wigner with an energy independent width convoluted with the experimental
resolution.  The experimental resolution in MeV/$c^2$ is approximately 
$\sigma(\textrm{MeV}/c^2) = -22.5 + 19.48m - 1.99m^2$ where $m$ is the mass in GeV/$c^2$.

\begin{figure}
\centerline{\includegraphics[width=5.5in]{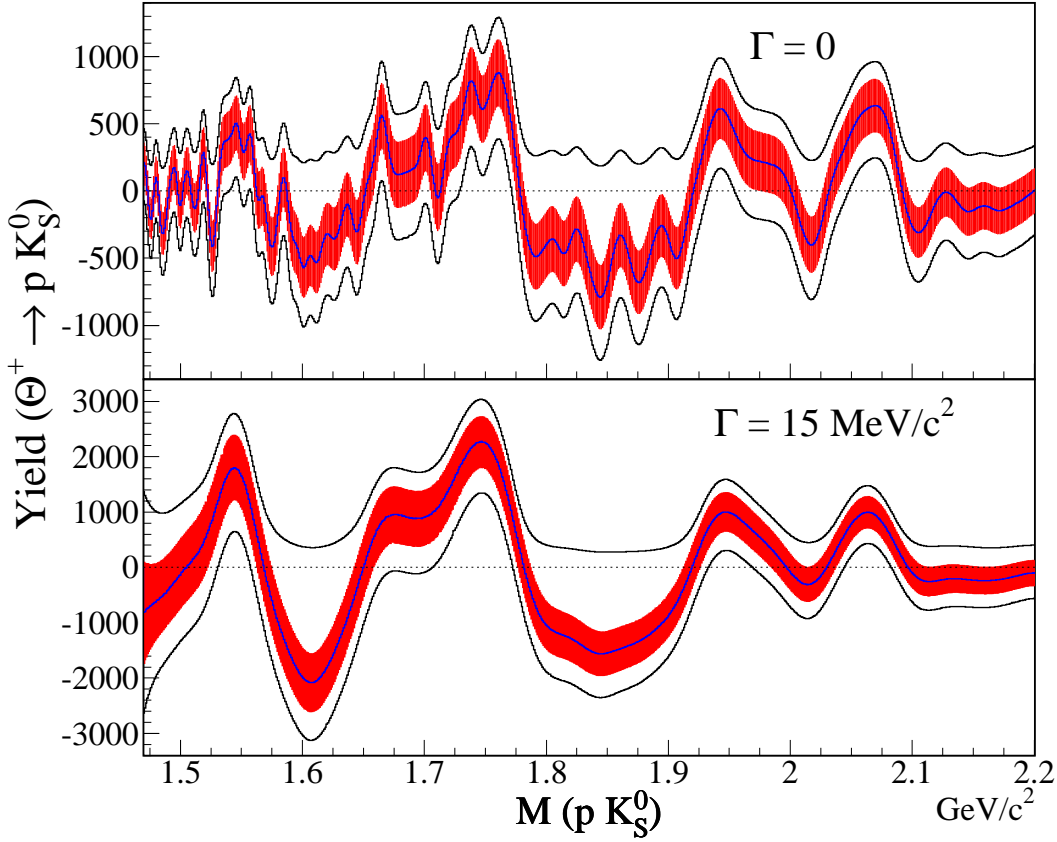}}
\caption{Pentaquark yields and upper limits.  Top (bottom) plots show
results for a natural width of $0$ ($15$~MeV/$c^2$).  The shaded region includes
the $1\:\sigma$ errors with the central value in the middle.  The outer curves show the upper and lower
limits.}
\label{fig:penta_yld}
\end{figure}

A series of 731 fits to the observed $pK_S^0$ mass plot were performed using the background and 
signal shapes described
above.  The signal mass is varied in 1~MeV/$c^2$ steps from $1470$ to $2200$~MeV/$c^2$ and
a binned log-likelihood fit using \textsc{Minuit}~\cite{minuit} is performed.
The $\pm1\:\sigma$ errors are defined as the point where $\Delta \log{\mathcal{L}} = 0.50$ relative to 
the maximum $\log{\mathcal{L}}$, while continually adjusting the background parameters to maximize 
$\log{\mathcal{L}}$.  
The 95\% CL 
lower limit is defined similarly with $\Delta \log{\mathcal{L}} = 1.92$. 
Both are obtained using \textsc{Minos}~\cite{minuit}.  The 95\% CL 
upper limit is constructed as follows:  
The likelihood function $\mathcal{L}$ versus yield is determined by maximizing $\log{\mathcal{L}}$ for 
many different (fixed) yields, allowing background parameters to float.  The likelihood
function is integrated from a yield of $0$ to $\infty$ to obtain the total likelihood.  The 95\% CL 
upper limit on the yield is defined as the point where 95\% 
of the total likelihood is between a yield of $0$ and the upper limit.  
This definition of an upper limit is used rather than a counting based Feldman--Cousins type limit due
to the large background which results in Gaussian errors.
The fitted yield, 1-$\sigma$ errors, and 95\% CL 
limits are shown in Fig.~\ref{fig:penta_yld}.  Of the 1462 fits, none of them finds a positive excursion
greater than 5$\sigma$.  Previous pentaquark observations have occurred between 1520~MeV/$c^2$ and 
1555~MeV/$c^2$ with a very small natural width.  In this region the largest excess we find is a 
2.5$\sigma$ excess at 1545~MeV/$c^2$ for the $\Gamma=0$~MeV/$c^2$ fits.  
A 3.1$\sigma$ excess is seen at the same location for the $\Gamma=15$~MeV/$c^2$
but this width would be inconsistent with the previous observations and is certainly not a 
convincing observation.  Additionally, this excess occurs in a region where the background 
distribution is peaking which can give the appearance of a signal.  Given the large number of fits, the
appearance of 2--3$\sigma$ excesses is not unlikely.

To compare with other experiments, the limits on yield must be converted to limits on production times 
(unknown) branching ratio.  We choose to normalize the $\Theta^+$ production 
cross section to $\Sigma^*(1385)^\pm$ and $K^*(892)^+$ because the reconstructed decay modes of these
particles $\Sigma^*(1385)^\pm \to \Lambda^0 \pi^\pm$ and $K^*(892)^+ \to K_S^0\pi^+$ are very similar,
in terms of topology and energy release, to the signal.  
Thus, we attempt to determine
\begin{equation}
\frac{\sigma\left(\Theta^+\right) \cdot \textrm{BR}\left(\Theta^+ \!\to\! p K_S^0\right)}{\sigma\left(K^*(892)^+\right)} \;\;\;\textrm{and}\;\;\;
\frac{\sigma\left(\Theta^+\right) \cdot \textrm{BR}\left(\Theta^+ \!\to\! p K_S^0\right)}{\sigma\left(\Sigma^*(1385)^\pm\right)}.
\label{eq:relx}
\end{equation}

Rewriting Eq.~\ref{eq:relx} in terms of measured yields (Y) and efficiencies ($\epsilon$), we find:
\begin{eqnarray}
\frac{\sigma\left(\Theta^+\right) \cdot \textrm{BR}\left(\Theta^+ \!\to\! p K_S^0\right)}{\sigma\left(K^*(892)^+\right)} &\;=\;&
\frac{\textrm{Y}\!\left(\Theta^+\right) \cdot \textrm{BR}\left(\Theta^+ \!\to\! p K_S^0\right)\cdot \epsilon_{K^*(892)^+}}
{\epsilon_{\Theta^+\!\to\! pK_S^0}\cdot \textrm{Y}\!\left(K^*(892)^+\right)} \notag \\
\frac{\sigma\left(\Theta^+\right) \cdot \textrm{BR}\left(\Theta^+ \!\to\! p K_S^0\right)}{\sigma\left(\Sigma^*(1385)^\pm\right)} &\;=\;&
\frac{\textrm{Y}\!\left(\Theta^+\right) \cdot \textrm{BR}\left(\Theta^+ \!\to\! p K_S^0\right)\cdot \epsilon_{\Sigma^*(1385)^\pm}}
{\epsilon_{\Theta^+\!\to\! pK_S^0}\cdot \textrm{Y}\!\left(\Sigma^*(1385)^\pm\right)}
\label{eq:relx2}
\end{eqnarray}

All of the efficiencies include the reconstruction and selection efficiencies plus corrections for unseen decays of
parent particles.  The $\Theta^+\!\to\! pK_S^0$ efficiency only includes the correction for the unseen $K_S^0$
decays, not corrections for $\Theta^+ \to pK_L^0$ or other $\Theta^+$ decay modes such as $\Theta^+\to nK^+$.
The $K^*(892)^+$ and $\Sigma^*(1385)$ efficiencies include all branching ratio corrections from the 
PDG~\cite{pdg}.  For 
$K^*(892)^+$ this corrections comes from BR$(K_S^0\to \pi^+\pi^-)=0.6861$, BR$(K^0\to K_S^0)=0.5$, and 
BR$(K^*(892)^+\to \overline{K}{}^0\pi^+)=0.667$ while for $\Sigma^*(1385)^\pm$ the relevant branching ratios are
BR$(\Lambda^0\to p\pi^-)=0.64$ and BR$(\Sigma^*(1385)^\pm \to \Lambda^0\pi^\pm)=0.88$.  Determining reconstruction
and selection efficiency (including acceptance) is described below.

The FOCUS detector is a forward spectrometer and therefore acceptance depends on
the produced particle momentum.  The production characteristics of the pentaquark are the
largest sources of systematic uncertainty in this analysis.  
We choose a particular production model to obtain limits and provide
sufficient information about the experiment for other interested parties to obtain limits based on other production 
models.  The production simulation begins with a library of $e^-$ and $e^+$ tracks obtained from a 
TURTLE simulation~\cite{turtle} of the Wideband beam line.  
From this library, an individual track is drawn and bremsstrahlung photons created by passage through 
a 20\% $X_0$ lead radiator.  Photons with energy above 15~GeV are passed to the \textsc{Pythia}~\cite{pythia} 
Monte Carlo simulation.  The \textsc{Pythia} simulation is run using minimum bias events 
(\texttt{MSEL=2}) with varying energies (\texttt{MSTP(171)=1}).
Options controlling parton distributions and gluon fragmentation were set to avoid heavy quark production 
(\texttt{MSTP(58)=3} and \texttt{MDME(156--160,1)=0}).  Also the center-of-mass minimum energy cut off was 
reduced from 10~GeV to 3~GeV (\texttt{PARP(2)=3}).  However, the minimum photon energy requirement gives an
effective minimum center-of-mass energy of 5.3~GeV.
Since \textsc{Pythia} does not produce pentaquarks, another particle must be
chosen to represent the pentaquark.  Other than mass, the most important effect on the production
is the number of quarks a particle has in common with the initially interacting hadrons, due to
the nature of the \textsc{Pythia} string fragmentation model.
The $\Xi^*(1530)^0$ and $\Sigma^*(1385)^+$ particles are chosen to represent the extremes in the
production of a pentaquark.  
The $\Xi^{*0}(ssu)$ $(\Sigma^+(suu))$ can obtain at most 33\% (67\%) 
of the remaining quarks from the target nucleon valence quarks, 
while the $\Theta^+(\overline{s}uudd)$ can take 60\%.  
In all cases,
the charge conjugate particles must obtain all quarks from the vacuum.  
The mass of the particle chosen to represent the
pentaquark, $\Xi^*(1530)^0$ or $\Sigma^*(1385)^+$, 
is set to the appropriate value in \textsc{Pythia}, 
by setting \texttt{PMAS(190,1)} or \texttt{PMAS(187,1)}, respectively.

\begin{figure}
\centerline{\includegraphics[width=4.5in,height=2.25in]{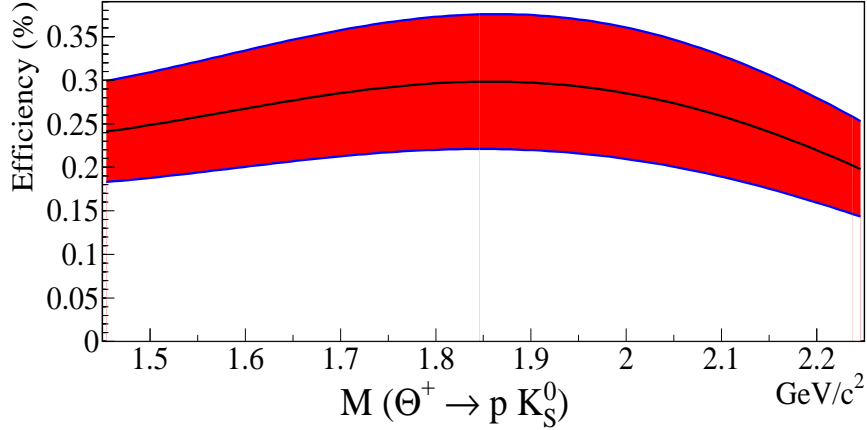}}
\caption{Acceptance versus mass for 
pentaquark candidates.  Upper (lower) curve is for a pentaquark produced as a
$\Xi^*(1530)^0$ $(\Sigma^*(1385)^+)$ and the middle is the average.}
\label{fig:accvsm}
\end{figure}

To calculate the relative cross sections in Eq.~\ref{eq:relx2} we need efficiencies for 
$\Sigma^*(1385)^\pm \to \Lambda^0\pi^\pm$, $K^*(892)^+ \to K_S^0\pi^+$, and $\Theta^+ \to p K_S^0$.  
These efficiencies are obtained from the FOCUS Monte Carlo simulation.  The 
dominant uncertainty in the efficiency determination is the modeling of the production 
characteristics of the parent particle.  For the observed particles, 
$\Sigma^*(1385)^\pm$ and $K^*(892)^+$, we can compare the data and Monte Carlo directly
and adjust the Monte Carlo simulation to produce the correct data distribution.  Even this is 
not sufficient, however, because areas where the efficiency is zero cannot be accounted for.
For $\Sigma^*(1385)^\pm$ and $K^*(892)^+$, we run a weighted Monte Carlo simulation which 
matches the Monte Carlo momentum distribution with the observed data momentum distribution in
the region for which the acceptance is not zero.  The dominant source of uncertainty for the
$\Sigma^*(1385)^\pm \to \Lambda \pi^\pm$ and $K^*(892)^+ \to K_S^0 \pi^+$ efficiencies is the
our lack of knowledge of the fraction of events completely outside of our acceptance (momentum less
than 15~GeV/$c$).  The weighted \textsc{Pythia} Monte Carlo predicts 67\% (79\%) of the
$K^*(892)^+$ $(\Sigma^*(1385)^\pm)$ particles are produced with momentum less than 15~GeV/$c$.
To obtain an estimate of the efficiency uncertainty, we assume that the number of 
particles with momentum less than 15~GeV/$c$ can be off by up to a factor of 2 (high or low).  
This leads to a relative
uncertainty on the $K^*(892)^+$ and $\Sigma^*(1385)^\pm$ efficiency of 43\% and 49\%, respectively.
The $\Theta^+\to pK_S^0$ efficiency is taken as the average of the efficiencies obtained from 
using $\Sigma^*(1385)^+$ and $\Xi^*(1530)^0$ as the substitute particle while the uncertainty is
half the difference between the two efficiencies.  The $\Theta^+\to pK_S^0$ with $K_S^0 \to \pi^+\pi^-$
efficiency versus mass (with no branching ratio corrections) is shown in Fig.~\ref{fig:accvsm}.
The average uncertainty in $\epsilon_{\Theta^+\to pK_S^0}$ is approximately 26\%.  It may seem incongruous
that the relative uncertainty of the efficiency of an unknown particle ($\sim$26\%) is less than that
for the high statistics normalizing modes ($>$40\%).  The efficiency uncertainty of the high 
statistics modes reflects the lack of knowledge of production outside of our acceptance.  However,
it is reasonable to assume that discrepancies in the Monte Carlo simulation will be similar for
the signal mode and the normalizing mode and therefore adding the uncertainty to the signal mode
is double-counting.  Note that the signal and normalizing efficiencies only appear as a ratio.

\begin{figure}
\centerline{\includegraphics[width=4.5in,height=2.25in]{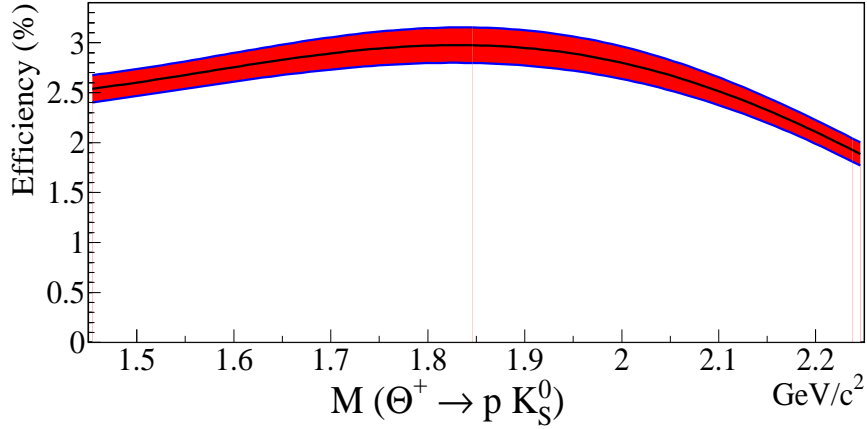}}
\caption{Acceptance versus mass for 
pentaquark candidates.  Lower (upper) curve is for a pentaquark produced as a
$\Xi^*(1530)^0$ $(\Sigma^*(1385)^+)$ and the middle is the average.  The pentaquark is
produced and reconstructed with momentum greater than 25~GeV/$c$.}
\label{fig:accvsma}
\end{figure}

We also report the relative cross sections in the region where our acceptance is good, that is for
parent particle momenta greater than 25~GeV/$c$.  This dramatically reduces the systematic uncertainties
associated with the measurement.  The uncertainty due to 
the production of $\Sigma^*(1385)^\pm$ and $K^*(892)^0$ is minimal.  The uncertainty in the
$\Theta^+$ efficiency is also dramatically reduced from approximately 26\% to about 6\% as shown in 
Fig.~\ref{fig:accvsma}.  The number of reconstructed $K^*(892)^0$, 
$\Sigma^*(1385)^+$, and $\Sigma^*(1385)^-$ at momenta greater than 25~GeV/$c$ is 7.88 million, 
127,000, and 212,000, respectively compared to the total sample of 8.98 million, 
151,000, and 256,000, respectively with no momentum cut.

The upper limit on the yield was obtained by mathematically integrating the likelihood function from 0 to 
$\infty$ and then integrating from 0 to 95\% of the total likelihood integral gave the 95\% CL upper limit.
To obtain the limit on cross section requires a different approach due to the significant systematic 
uncertainties.  We use a method based on a note by Convery~\cite{convery} which is inspired by the Cousins and 
Highland~\cite{cousins} philosophy for including systematic uncertainties.  The Cousin and Highland 
prescription is appropriate for low background experiments with Poisson errors while the Convery proposal
is applicable to the Gaussian errors which result from the large background in our case.
Modifications to the Convery approach are made to give an exact solution~\cite{stenson}.  

Before systematics are considered,
an analysis using a maximum likelihood fit returns a central value for the branching ratio $(\hat{B})$
and a statistical error $(\sigma_B)$.  The likelihood function is
\begin{equation}
p(B) \propto \exp{\!\left[\frac{-(B-\hat{B})^2}{2 \sigma_B^2}\right]}
\end{equation}
Following the notation in Convery, we associate $\hat{S}$ with the nominal efficiency and $\sigma_S$ as
the error on the efficiency.  Adding the uncertainty on the efficiency changes the likelihood to:
\begin{equation}
\label{eq:integral}
p(B) \propto \int_0^1 \exp{\!\left[\frac{-(S B/\hat{S} - \hat{B})^2}{2 \sigma_B^2}\right]} 
\exp{\!\left[\frac{-(S-\hat{S})^2}{2\sigma_s^2}\right]} d\!S
\end{equation}
Using Mathematica$^\textrm{\textregistered}$, removing unimportant multiplicative constants, and changing variables from
 $\sigma_S$ to $\sigma_\epsilon \equiv \sigma_S/\hat{S}$, the integral in Eq.~\ref{eq:integral} becomes:
\begin{multline}
\label{eq:fulleps}
p(B) \propto \frac{1}{\sqrt{\frac{B^2}{\sigma_B^2}+\frac{1}{\sigma_\epsilon^2}}}
\exp{\!\left[\frac{-(B-\hat{B})^2}{2(B^2\sigma_\epsilon^2+\sigma_B^2)}\right]}\left\{
\textrm{erf}\!\left[\frac{B \hat{B} \sigma_\epsilon^2 + \sigma_B^2}{\sqrt{2}\sigma_\epsilon \sigma_B \sqrt{B^2\sigma_\epsilon^2 + \sigma_B^2}}\right]\right. - \\
\left. \textrm{erf}\!\left[\frac{(\hat{S}-1)\sigma_B^2 - B\sigma_\epsilon^2 (B - \hat{B}\hat{S})}{\sqrt{2}\hat{S}\sigma_\epsilon \sigma_B \sqrt{B^2 \sigma_\epsilon^2 + \sigma_B^2}}\right]\right\}
\end{multline}

We integrate Eq.~\ref{eq:fulleps} from 0 to $\infty$ to obtain the total probability and then integrate from 0 
to the point at which 95\% of the total probability is included and define this as our 95\% CL upper limit.
The branching ratio $B$ of Eq.~\ref{eq:fulleps} is simply the relative cross sections times the unknown 
pentaquark branching ratio as in Eq.~\ref{eq:relx2}.  The relative uncertainties on the efficiency for the 
signal and normalizing mode are added in quadrature to become $\sigma_\epsilon$ in Eq.~\ref{eq:fulleps}.
Furthermore, $\hat{S}$ is the relative efficiency between the signal and normalizing modes and $\sigma_B$ is
the statistical uncertainty on the branching ratio due simply to the uncertainty in the signal yield.

\begin{figure}
\centerline{\includegraphics[width=5.0in]{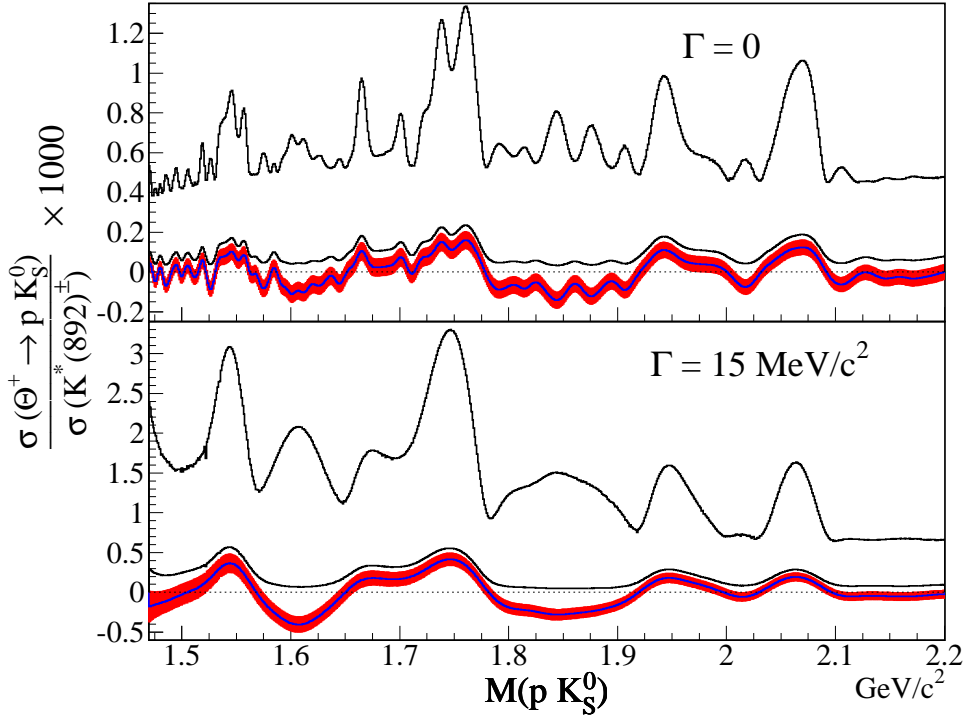}}
\caption[$\frac{\sigma(\Theta^+)\times \textrm{BR}(\Theta^+\to pK_S^0)}{\sigma(K^*(892)^+)}$ versus mass]
{$\frac{\sigma(\Theta^+)\times \textrm{BR}(\Theta^+\to pK_S^0)}{\sigma(K^*(892)^+)}$ versus mass.
Top (bottom) plots show results for a $\Theta^+$ natural width of $0$ ($15$~MeV/$c^2$).  The shaded region encompasses
the $1\:\sigma$ statistical uncertainty with the central value in the middle.  The top curve shows the 95\% CL upper limit
including systematic uncertainties while the middle curve is the 95\% CL upper limit with statistical uncertainties only.}
\label{fig:xsecul_kstar}
\end{figure}

\begin{figure}
\centerline{\includegraphics[width=5.0in]{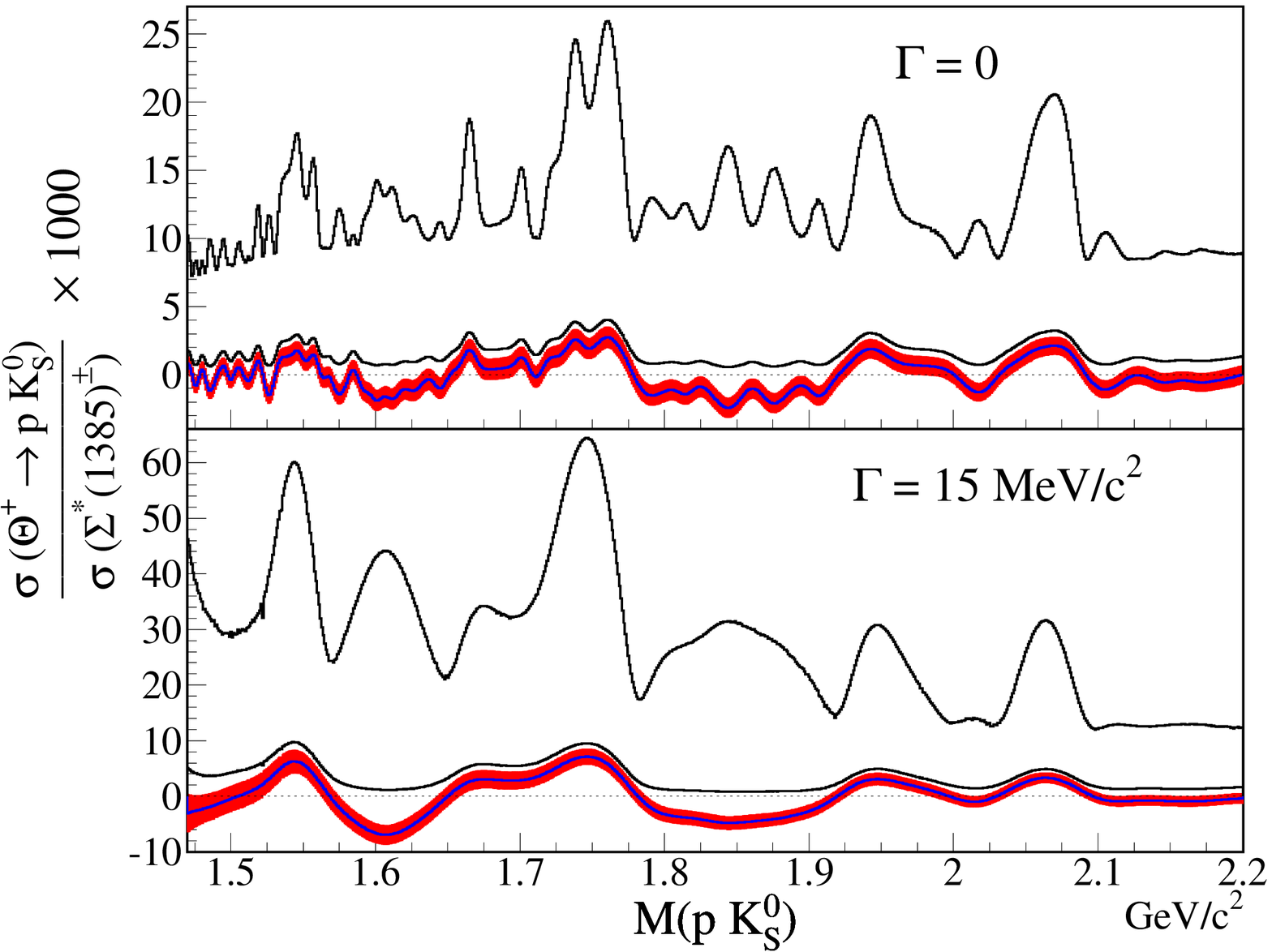}}
\caption[$\frac{\sigma(\Theta^+)\times \textrm{BR}(\Theta^+\to pK_S^0)}{\sigma(\Sigma^*(1385)^+) + \sigma(\Sigma^*(1385)^-)}$ versus mass]
{$\frac{\sigma(\Theta^+)\times \textrm{BR}(\Theta^+\to pK_S^0)}{\sigma(\Sigma^*(1385)^+) + \sigma(\Sigma^*(1385)^-)}$ versus mass.
Top (bottom) plots show results for a $\Theta^+$ natural width of $0$ ($15$~MeV/$c^2$).  The shaded region encompasses
the $1\:\sigma$ statistical uncertainty with the central value in the middle.  The top curve shows the 95\% CL upper limit
including systematic uncertainties while the middle curve is the 95\% CL upper limit with statistical uncertainties only.}
\label{fig:xsecul_sigma}
\end{figure}

Figure~\ref{fig:xsecul_kstar} shows the results for 
$\frac{\sigma\left(\Theta^+\right) \cdot \textrm{BR}\left(\Theta^+ \!\to\! p K_S^0\right)}{\sigma\left(K^*(892)^+\right)}$
with an assumed natural width of 0 (15)~MeV/$c^2$ for the top (bottom) plot.  This is the result corrected for all undetected
particles.
The shaded band shows the $\pm1\sigma$ limits with statistical uncertainties only; the line in the middle of the band is the
central value.  The top curve shows the 95\% CL upper limit using the method described above including statistical and
systematic uncertainties.  The curve between the full upper limit and the $1\sigma$ band is the 95\% CL upper limit using
the method described above with no systematic uncertainties included.  The large systematic uncertainties are due to the
attempt to correct for the vast majority of particles outside of our acceptance.  While this systematic uncertainty 
significantly degrades
the limit, the production times branching ratio of the pentaquark relative to $K^*(892)^+$ production is still less than 
$0.0013$ $(0.0032)$ (95\% CL) over the entire mass range for a natural width of 0 (15)~MeV/$c^2$.  
Figure~\ref{fig:xsecul_sigma} gives the same results for 
$\frac{\sigma\left(\Theta^+\right) \cdot \textrm{BR}\left(\Theta^+ \!\to\! p K_S^0\right)}{\sigma\left(\Sigma^*(1385)^\pm\right)}$.
In this case, the 95\% CL upper limit on the pentaquark production times branching ratio relative to $\Sigma^*(1385)^\pm$ is
$0.025$ $(0.062)$ over the entire mass range for a natural width of 0 (15)~MeV/$c^2$.

\begin{figure}
\centerline{\includegraphics[width=5.0in]{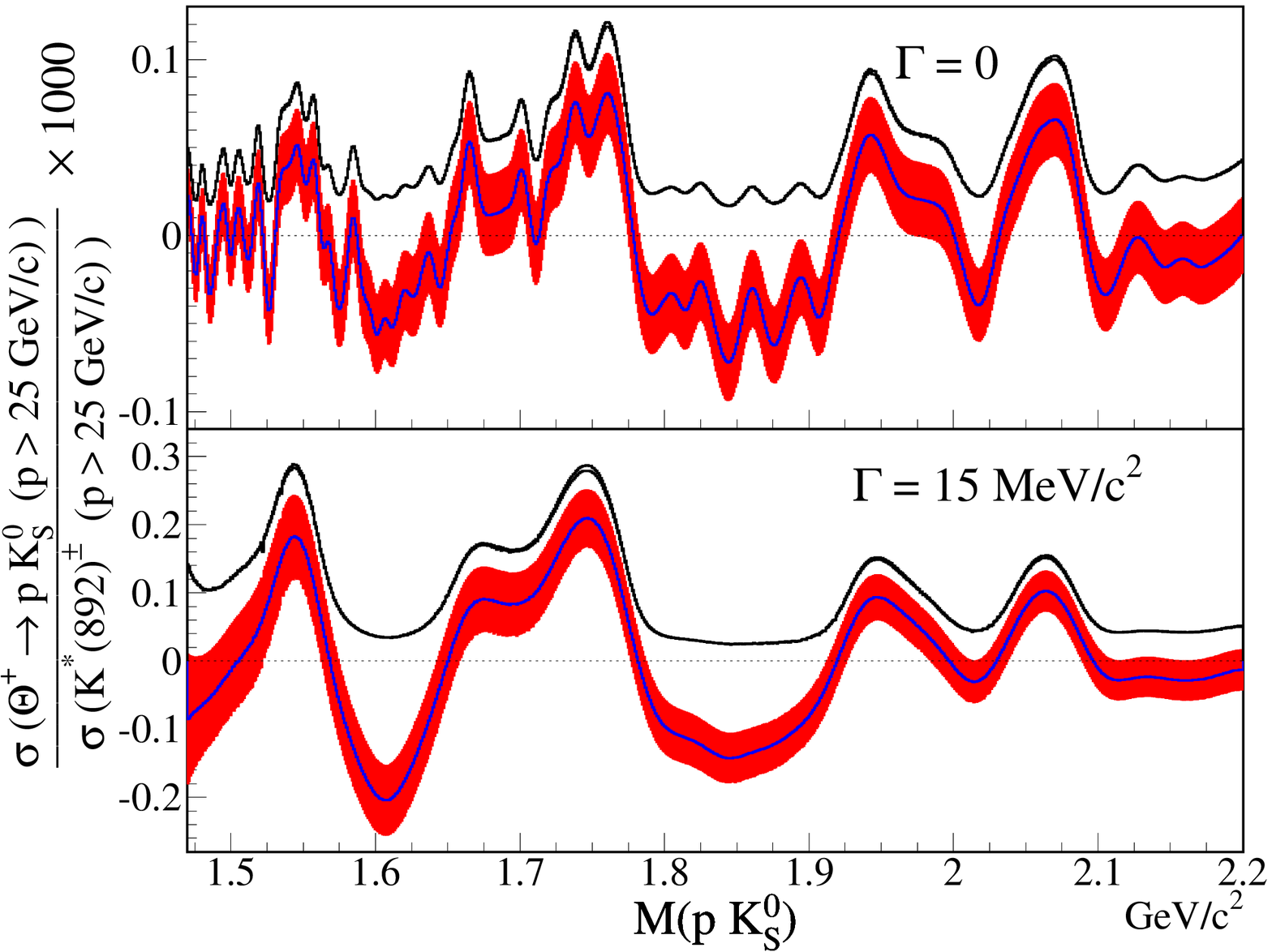}}
\caption[$\frac{\sigma(\Theta^+)\times \textrm{BR}(\Theta^+\to p K_S^0)}{\sigma(K^*(892)^+)}$ for $p>25$~GeV/$c$ versus mass]
{$\frac{\sigma(\Theta^+)\times \textrm{BR}(\Theta^+\to pK_S^0)}{\sigma(K^*(892)^+)}$ for $p>25$~GeV/$c$ versus mass.
Top (bottom) plots show results for a $\Theta^+$ natural width of $0$ ($15$~MeV/$c^2$).  The shaded region encompasses
the $1\:\sigma$ statistical uncertainty with the central value in the middle.  The top curve shows the 95\% CL upper limit
including systematic uncertainties and is virtually indistinguishable from the middle curve which shows the 95\% CL upper limit with 
statistical uncertainties only.}
\label{fig:xsecula_kstar}
\end{figure}

\begin{figure}
\centerline{\includegraphics[width=5.0in]{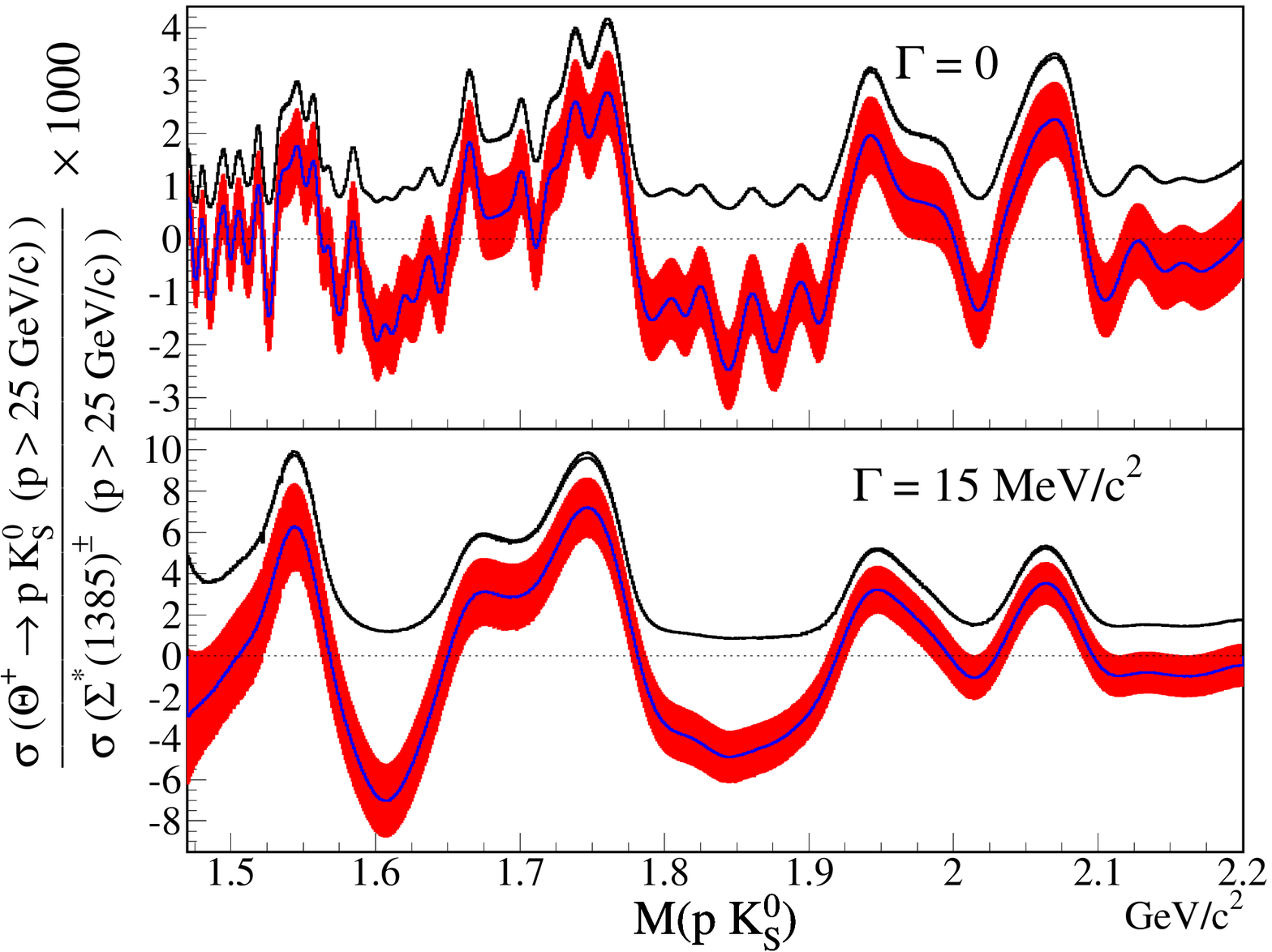}}
\caption[$\frac{\sigma(\Theta^+)\times \textrm{BR}(\Theta^+\to pK_S^0)}{\sigma(\Sigma^*(1385)^+) + \sigma(\Sigma^*(1385)^-)}$ for $p>25$~GeV/$c$ versus mass]
{$\frac{\sigma(\Theta^+)\times \textrm{BR}(\Theta^+\to pK_S^0)}{\sigma(\Sigma^*(1385)^+) + \sigma(\Sigma^*(1385)^-)}$ for $p>25$~GeV/$c$ versus mass.
Top (bottom) plots show results for a $\Theta^+$ natural width of $0$ ($15$~MeV/$c^2$).  The shaded region encompasses
the $1\:\sigma$ statistical uncertainty with the central value in the middle.  The top curve shows the 95\% CL upper limit
including systematic uncertainties and is virtually indistinguishable from the middle curve which shows the 95\% CL upper limit 
with statistical uncertainties only.}
\label{fig:xsecula_sigma}
\end{figure}

Figures~\ref{fig:xsecula_kstar} and \ref{fig:xsecula_sigma} show the same results for the restricted range of momentum greater
than 25 GeV/$c$.  That is, they show limits on relative cross sections for particles ($\Theta^+$, $K^*(892)^+$, 
$\Sigma^*(1385)^\pm$) produced with $p > 25$~GeV/$c$.

\section{Conclusions}

We find no evidence for pentaquarks decaying to $p K_S^0$ in the mass range of
1470~MeV/$c^2$ to 2200~MeV/$c^2$.  In contrast, we observe 9 million $K^*(892)^+ \to K_S^0\pi^+$ 
particles and 0.4 million $\Sigma^*(1385)^\pm \to \Lambda^0 \pi^\pm$ particles which have
a very similar topology and energy release.
We set 95\% CL upper limits on the yield over the entire mass range with a maximum of 1300 (3000) 
events for an assumed natural width of 0 (15)~MeV/$c^2$.  We also 
obtain 95\% CL upper limits on the cross section for pentaquark production times the branching ratio to 
$pK_S^0$ relative to $K^*(892)^+ \to K_S^0\pi^+$ and $\Sigma^*(1385)^\pm \to \Lambda^0 \pi^\pm$.  
These limits are determined for two cases.  The first case is for parent particles produced at any momenta 
(albeit with a minimum center-of-mass energy of 5.3~GeV) where we find a maximum upper limit of 
$\frac{\sigma\left(\Theta^+\right) \cdot \textrm{BR}\left(\Theta^+ \!\to\! p K_S^0\right)}{\sigma\left(K^*(892)^+\right)} < 0.0013\; (0.0033)$
and $\frac{\sigma\left(\Theta^+\right) \cdot \textrm{BR}\left(\Theta^+ \!\to\! p K_S^0\right)}{\sigma\left(\Sigma^*(1385)^\pm\right)} < 0.023\; (0.057)$
at 95\% CL for a natural width of 0 (15)~MeV/$c^2$.
In the second case we measure the relative cross sections for parent particles with momenta above 25~GeV/$c$ 
(a region of good acceptance) and calculate 95\% CL limits of 
$\frac{\sigma\left(\Theta^+\right) \cdot \textrm{BR}\left(\Theta^+ \!\to\! p K_S^0\right)}{\sigma\left(K^*(892)^+\right)} < 0.00012\; (0.00029)$
and $\frac{\sigma\left(\Theta^+\right) \cdot \textrm{BR}\left(\Theta^+ \!\to\! p K_S^0\right)}{\sigma\left(\Sigma^*(1385)^\pm\right)} < 0.0042\; (0.0099)$
for a natural width of 0 (15)~MeV/$c^2$.  

Very few of the observing experiments report results for $K^*(892)^+ \to K_S^0\pi^+$ and
$\Sigma^*(1385)^\pm \to \Lambda^0 \pi^\pm$ yields.  One CLAS result apparently finds a yield of
$\sim$1000 $K^*(892)^+$~\cite{clas_back} while SVD reconstructs $\sim$125 $K^*(892)^+ \to K_S^0\pi^+$ and 
$\sim$100 $\Sigma^*(1385)^+\to \Lambda^0\pi^+$ decays~\cite{svd}.  The 
FOCUS results presented here represent samples that are more than 2000 times larger.  Unfortunately,
differences in production between (mostly) low energy experiments which have reported observations and
a high energy experiment such as FOCUS prevent any definitive conclusions from being drawn.

\section{Acknowledgments}
We wish to acknowledge the assistance of the staffs of Fermi National
Accelerator Laboratory, the INFN of Italy, and the physics departments
of the collaborating institutions. This research was supported in part
by the U.~S.  National Science Foundation, the U.~S. Department of
Energy, the Italian Istituto Nazionale di Fisica Nucleare and
Ministero dell'Istruzione dell'Universit\`a e della Ricerca, the
Brazilian Conselho Nacional de Desenvolvimento Cient\'{\i}fico e
Tecnol\'ogico, CONACyT-M\'exico, the Korean Ministry of Education, 
and the Korean Science and Engineering Foundation.

\bibliographystyle{unsrt}

\end{document}